\documentclass[10pt,final,journal,letterpaper,twoside,twocolumn]{IEEEtran}
\usepackage{cite}
\usepackage[cmex10]{amsmath}
\usepackage{placeins}
\usepackage{graphicx,epsfig,amssymb}
\usepackage{anysize}
\marginsize{0.8in}{0.8in}{0.25in}{0.25in} 

\begin{document}

\title{Spectrum and Energy Efficiency Evaluation of Two-Tier Femtocell networks With Partially Open Channels}
\author{Xiaohu Ge,~\IEEEmembership{Senior~Member,~IEEE,}
        Tao Han, ~\IEEEmembership{Member,~IEEE,}
        Yan Zhang,~\IEEEmembership{Senior~Member,~IEEE,}
        Guoqiang Mao,~\IEEEmembership{Senior~Member,~IEEE,}
        Cheng-Xiang Wang,~\IEEEmembership{Senior~Member,~IEEE,}
        Jing Zhang,~\IEEEmembership{Member,~IEEE,}
        Bin Yang,
        and Sheng Pan
\thanks{\scriptsize{Copyright (c) 2013 IEEE. Personal use of this material is permitted. However, permission to use this material for any other purposes must be obtained from the IEEE by sending a request to pubs-permissions@ieee.org.}}
\thanks{\scriptsize{X. Ge, T. Han (corresponding author), J. Zhang, B. Yang and S. Pan are with the Department of Electronics and Information Engineering, Huazhong University of Science and Technology, Wuhan 430074, Hubei, China (email: \{xhge, hantao, zhangjing, yangbin, pansheng\}@mail.hust.edu.cn).}}
\thanks{\scriptsize{Y. Zhang is with Simula Research Laboratory, 1364 Fornebu, Norway. (email:yanzhang@ieee.org).}}
\thanks{\scriptsize{G. Mao is with School of Computing and Communications,
University of Technology, Sydney and National ICT Australia, Sydney, Australia. (email: g.mao@ieee.org).}}
\thanks{\scriptsize{C.-X. Wang is with Joint Research Institute for Signal and Image Processing, School of Engineering \& Physical Sciences, Heriot-Watt University, Edinburgh, EH14 4AS, UK. (email: cheng-xiang.wang@hw.ac.uk).}}
\thanks{\scriptsize{The authors would like to acknowledge the support from the National Natural Science Foundation of China (NSFC) under the grants 60872007 and 61271224, NFSC Major International Joint Research Project under the grant 61210002, the Ministry of Science and Technology (MOST) of China under the grants 0903, the Hubei Provincial Science and Technology Department under the grant 2011BFA004 and 2013BHE005, and the Fundamental Research Funds for the Central Universities under the grant 2011QN020 and 2013ZZGH009. This research is partially supported by RCUK for the UK-China Science Bridges Project: R\&D on (B)4G Wireless Mobile Communications (EP/G042713/1), the European Commission FP7 Project EVANS (grant no. 2010-269323), SmartGrids ERA-Net project PRO-NET funded through Research Council of Norway (project 217006), EU FP7-PEOPLE-IRSES, project acronym S2EuNet (grant no. 247083), project acronym WiNDOW (grant no. 318992) and the Key Laboratory of Cognitive Radio and Information Processing (Guilin University of Electronic Technology), the Ministry of Education, China (Grant No.: 2013KF01). Guoqiang Mao's work is supported by Australian Research Council Discovery projects DP110100538 and DP120102030.
}}
}
\maketitle

\markboth{IEEE TRANSACTIONS on VEHICULAR TECHNOLOGY, Vol. XX, No. Y, Month 2014} {Ge \MakeLowercase{\textit{et al.}}: Spectrum and Energy Efficiency Evaluation of Two-Tier Femtocell networks With Partially Open Channels}%

\begin{abstract}
Two-tier femtocell networks is an efficient communication architecture that
significantly improves throughput in indoor environments with low power
consumption. Traditionally, a femtocell network is usually configured to be
either completely open or completely closed in that its channels are either made available to all users or used by its own users only. This may limit network
flexibility and performance. It is desirable for owners of femtocell base stations if a femtocell can
partially open its channels for external users access. In such scenarios, spectrum and
energy efficiency becomes a critical issue in the design of femtocell network
protocols and structure. In this paper, we conduct performance analysis for
two-tier femtocell networks with partially open channels. In particular, we
build a Markov chain to model the channel access in the femtocell network and then derive the performance
metrics in terms of the blocking probabilities. Based on stationary state probabilities derived by Markov chain models, spectrum and energy efficiency are modeled and analyzed under different
scenarios characterized by critical parameters, including number of femtocells in a macrocell, average number of users, and number of open channels in a femtocell. Numerical and Monte-Carlo (MC) simulation results indicate that the number of open channels in a femtocell has an adverse
impact on the spectrum and energy efficiency of two-tier femtocell networks. Results in
this paper provide guidelines for trading off spectrum and energy efficiency of two-tier femtocell networks by configuring different numbers of open channels in a femtocell.
\end{abstract}


\section{Introduction}
\IEEEPARstart{T}{o} accommodate the rapid increase of wireless data traffic in indoor environments, two-tier femtocell networks have been proposed to support the high spectrum efficiency in wireless communications \cite{Chandrasekhar08,Golaup09}. Particularly, energy efficiency in wireless communications has received lots of attention lately \cite{Humar11,Hasan11}. It is important to study the spectrum and energy efficiency of femtocell networks for both economical and environmental considerations \cite{Chen11}.

There are several studies on spectrum efficiency of femtocell networks in the literature \cite{Kim10,Elkourdi11,Zhang10,Pantisano12,Cheung12,Xiang10,Oh11,Chandrasekharand09,Sun12,Chandrasekhar09,Jo09,Xia10}. They mainly focus on evaluating the impact of interference on the capacity and frequency reuse in wireless femtocell networks. Kim \textit{et al.} derived the per-tier outage probability, i.e., the macrocell outage probability and the femtocell outage probability, by using a simplified mathematical model to closely approximate the femtocell interference distribution in a two-tier femtocell network \cite{Kim10}. Then, the capacity of the co-channel two-tier networks with outage probability constraints was obtained. Elkourdi and Simeone proposed a new approach to enable cooperations between femtocell base stations and macrocell base stations \cite{Elkourdi11}. In addition, the tradeoff between the outage probability and the diversity-multiplexing gains for both uplinks and downlinks was evaluated. Zhang focused on studying the blocking probability of femtocell networks assuming completely closed channel access. The paper recommended the use of a small number of split spectrum in a femtocell to increase the service availability in a macrocell \cite{Zhang10}. Also Pantisano \textit{et al.} \cite{Pantisano12} adopt a closed access scheme at each femtocell to study the proposed novel framework of cooperation among femtocell users and macrocell users. Results have shown that the performance of femtocell users and macrocell users are respectively limited by interference and delay. Based on stochastic geometric techniques, the transmission success probability is derived for two cases of closed access and open access schemes at the femtocell, respectively \cite{Cheung12}. Xiang \textit{et al.} applied the cognitive radio technology in femtocell
networks and formulated the downlink spectrum sharing problem as a mixed integer nonlinear programming problem \cite{Xiang10}. A joint channel allocation and fast power control scheme was proposed to improve the spectrum efficiency of femtocell networks \cite{Oh11}. Chandrasekhar and Andrews analyzed the effect of channel uncertainty on two-tier femtocell networks. The transmit power level was determined to provide the desired signal-to-interference-plus-noise ratio (SINR) for the indoor cell edge femtocell users \cite{Chandrasekharand09}. On that basis, the beam weight was further optimized to maximize the output SINR of macrocell users and femtocell users. Sun \textit{et al.} proposed an inter- and intra-tier interference mitigation strategy and applied the strategy to a partial co-channel assignment problem \cite{Sun12}. In their proposed scheme, macrocell users are divided into femto-interfering users and regular users, and then an auction-based subcarrier allocation algorithm was developed for mitigating intra-tier interference and improving the spectrum efficiency in femtocell networks. Chandrasekhar \textit{et al.} developed a link quality protection algorithm for progressively reducing the SINR targets when a cellular user is unable to meet its SINR target in a two-tier femtocell network \cite{Chandrasekhar09}. Jo \textit{et al.} discovered that both the open-loop and the closed-loop control schemes can effectively compensate the uplink throughput degradation of the macrocell
base station in a two-tier femtocell network \cite{Jo09}. Xia\textit{ et al.} evaluated both the completely open and completely closed
femtocell access schemes using theoretical analysis and simulations for code division multiple access (CDMA), time division multiple access (TDMA) and orthogonal frequency division multiple access (OFDMA) scenarios \cite{Xia10}.

Considering the low power advantage of femtocells, energy efficiency in
femtocell networks has attracted attention in recent studies \cite{Hou10,Khirallah11,Mclaughlin11,Zhang11,Cao10,Domenico11,Apio11,Ku13,Hong13}. Particularly, Hou
and Laurenson showed that the cellular and femtocell heterogeneous
network architecture is able to provide a high quality of service (QoS)
while significantly reduce power consumption \cite{Hou10}. Khirallah \textit{et
al.} proposed an approach to estimate the total energy consumption in homogeneous and
heterogeneous networks with femtocell deployments \cite{Khirallah11}. Mclaughlin \textit{et al.}
proposed a power allocation strategy to optimize both spectral and energy efficiency for large-scale femtocells deployment \cite{Mclaughlin11}. Zhang \textit{et al.} reported
various power control and radio resource management schemes for long term
evolution advanced (LTE-A) networks employing femtocells \cite{Zhang11}. Their results demonstrated
that the femtocell provides an energy efficient solution for indoor coverage in LTE-A
networks. Cao and Fan validated the energy efficiency improvement by simulating a LTE
femtocell networks with realistic system parameters \cite{Cao10}. Domenico \textit{et al.}
proposed two radio resource management schemes to enhance the energy
efficiency of two-tier femtocell networks while improving both macrocell and
femtocell throughput \cite{Domenico11}. Apio \textit{et al.} presented a switch-off
algorithm to reduce the power consumption of some stations during low traffic periods in
two-tier femtocell networks \cite{Apio11}. Ku \textit{et al.} explored the tradeoff between the spectrum efficiency and energy efficiency in wireless networks \cite{Ku13}. Hong \textit{et al.} evaluated the Energy-spectrum efficiency tradeoff in virtual MIMO systems \cite{Hong13}.

In this paper, we study both spectrum efficiency and energy efficiency in
a two-tier femtocell network. Different from traditional studies where a femtocell is configured to be either completely open or completely
closed, we allow a subset of channels to be open while the other channels to be closed in a femtocell. In particular, some femtocell channels are open for all users and the rest of femtocell
channels can only be used by the femtocell's own customers. We call this channel arrangement \textit{partially open channel arrangement}. This partially open channel arrangement is shown to be able to
significantly improve network flexibility and satisfy different requirements from femtocell owners. Moreover, this \textit{partially open channel arrangement} is valuable for trading off the spectrum and energy efficiency of two-tier femtocell networks by changing number of open channels in a femtocell. We also conduct performance analysis for two-tier femtocell networks with partially open channels based on Markov chain models.
Such analysis is important to quantitatively characterize the spectrum and energy
efficiency in a two-tier femtocell network. Specifically the major contributions of this paper
are:

\begin{enumerate}
\item A Markov chain model is presented for two-tier femtocell networks with partially
open channels. Furthermore, based on the stationary state probabilities of equilibrium equations in the Markov chain model, the closed-form blocking probability models of femtocell users and macrocell users are derived and analyzed.
\item Spectrum and energy efficiency models are proposed for two-tier femtocell networks with partially open channels based on the Markov chain model. Analytical results of spectrum and energy efficiency models provide guidelines for trading off the spectrum and energy efficiency of two-tier femtocell networks by configuring different numbers of open channels in a femtocell.
\item Numerical and Monte-Carlo (MC) simulation results are compared to validate the accuracy of the analysis. Moreover, simulation results demonstrate the tradeoff between spectrum efficiency and energy efficiency in two-tier femtocell networks with partially open channels.
\end{enumerate}

The rest of the paper is organized as follows. Section~\ref{sec2} describes the system model of a
two-tier femtocell network with partially open channels. In Section~\ref{sec3}, a
Markov chain state transition diagram is proposed to model the
dynamics of a femtocell. Based on the Markov chain model, the spectrum efficiency and the energy efficiency are investigated for femtocell networks with partially
open channels in Section~\ref{sec4}. Section~\ref{sec5} is the numerical results and discussions of spectrum and energy efficiency. Finally, Section~\ref{sec6} concludes this paper.

\section{Two-Tier Femtocell network System Model}
\label{sec2}

Fig. 1 shows the system model of a two-tier femtocell network. A macrocell
network consists of multiple macrocells that share the same bandwidth. A
macrocell covers a regular circular region ${A_M}$ with radius ${R_M}$  and a
macrocell base station (BS) is located in the center of the circle. The macrocell BS has
${N_M}$ channels which are open for all users. A femtocell covers a circular region ${A_F}$  with radius ${R_F}$ and a femtocell BS, which is usually called as an access point (AP), is located in the center of the femtocell. A femtocell BS has ${N_F}$ channels
which are classified into two types: one type of channels is open in the
sense that they can be used by all users; the other type of channels is
called closed channels which can only be used by femtocell users. The number
of open femtocell channels is denoted by ${N_{F\_O}}$ and the number of
closed femtocell channels then becomes ${N_F} - {N_{F\_O}}$ in a femtocll.

\begin{figure}
\vspace{0.1in}
\centerline{\includegraphics[width=8cm,draft=false]{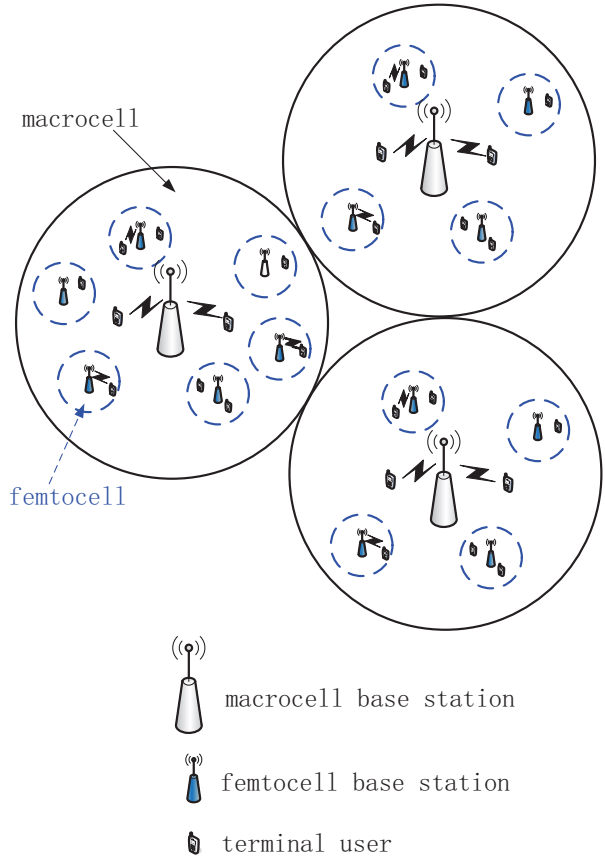}}
\caption{\small Two-tier femtocell network system model.}
\end{figure}

We consider two kinds of users in the two-tier femtocell networks: macrocell
users and femtocell users. Macrocell users can access all unoccupied
macrocell channels and unoccupied open femtocell channels when macrocell users are located in corresponding coverage areas of these femtocells. Femtocell users can
access all unoccupied macrocell channels and unoccupied femtocell
channels in which these femotocell users are located. Within a specific femtocell, a femtocell user accesses the channels in the following order: a femtocell user will first access an unoccupied closed femtocell channel if there exists an unoccupied closed femtocell channel in the specific femtocell; a femtocell user will access an unoccupied open femtocell channel if all closed femtocell channels are busy and there exists an unoccupied open femtocell channel in the specific femtocell. Finally, the femtocell user will access an unoccupied macrocell channel if all closed femtocell channels and open femtocell channels are busy in the specific femtocell. That is, a femtocell user can only access an unoccupied open femtocell channel when all closed femtocell channels are busy, and similarly for unoccupied macrocell channels. The macrocell network is overlaid with a femtocell network consisting
of multiple femtocells. Femtocells do not overlap with each other.
Therefore, users can hand off between a macrocell and a femtocell or between
a macrocell and an adjacent macrocell. It is assumed that there are a total number of N femtocell BSs uniformly distributed within a macrocell and futhermore there are an average number of M femtocell users uniformly distributed in a femtocell. Without loss of generality, it is assumed that the traffic process originating from a user is governed by a Poisson process. Consequently, both the inter-traffic arrival time and the traffic duration follow exponential distributions. To facilitate reading, the notations and symbols used in this paper are listed in TABLE I.

\begin{table*}[htbp]
\centering
\caption{NOTATIONS AND SYMBOLS USED IN THE PAPER}
\begin{tabular}{l|l}
\hline Symbol & Definition/explanation \\
\hline
${A_M}$, ${A_F}$ & The areas of a macrocell and a femtocell, respectively \\
${B_w}$ & The bandwidth of femtocell channel \\
$G$ & The standard random variable with a normal distribution \\
${I_k}$ & The interference power received from the ${k^{th}}$ adjacent femtocell \\
${L}$ & The distance between the user and the corresponding femtocell BS \\
$l$ & The distance between two femtocell BSs in a macrocell \\
${M}$ & The average number of femtocell users in a femtocell \\
${N}$ & The total number of femtocells within a macrocell \\
${N_F}$, ${N_M}$ & The number of channels in a femtocell and a macrocell, respectively \\
${N_{F\_O}}$ & The number of open femtocell channels in a femtocell \\
${n_0}$ & The additive white Gaussian noise \\
${n_w}$ & The number of walls among femtocells in an indoor environment \\
${P_{closed}}$, ${P_{open}}$ & The occupancy probability of a closed femtocell channel and an open femtocell channel, respectively \\
${P_{FU\_F}}$, ${P_{MU\_F}}$ & The blocking probability of femtocell user and macrocell user in a femtocell, respectively\\
${P_{Handoff\_MF}}$ & The hand-off probability from a macrocell to a femtocell \\
${P_{Handoff\_FM}}$ & The hand-off probability from a femtocell to a macrocell \\
${P_{Handoff\_MM}}$ & The hand-off probability from a macrocell to one of adjacent macrocells \\
${P_{U\_M}}$ & The user blocking probability in a macrocell\\
${PW_{FBS}}$ & The total energy consumption of femtocell BS \\
${PW_c}$, ${PW_t}$ & The fixed and dynamic energy consumption of femtocell BS, respectively \\
${PW_v}$ & The transmission power over a femtocell channel \\
$q$ & The fraction of femtocell calls to the total number of calls originating from a femtocell user \\
${R_F}$, ${R_M}$ & The radius of a femtocell and a macrocell, respectively \\
${R_p}$ & the protection distance between the user and the corresponding femtocell BS \\
${S_m}$ & The user desired signal power \\
${T_{cF\_F}}$, ${T_{cM\_F}}$ & The channel holding time of femtocell users and macrocell users in a femtocell, respectively \\
${T_{cF\_M}}$, ${T_{cM\_M}}$ & The channel holding time of femtocell user and macrocell user in a macrocell, respectively \\
${T_F}$, ${T_M}$ & The traffic intensity of femtocell users and macrocell users in a macrocell, respectively \\
${Z_{shadowing}}$ & The shadowing effect in an indoor environment \\
${\alpha ^2}$ & The random variable is exponentially distributed with mean value 1 in a Rayleigh fading environment \\
$\beta$ & The path loss exponent over femtocell channels \\
${\eta _{EE}}$ & The utility function of energy efficiency \\
${\lambda _1}$, ${\lambda _2}$ & The aggregate traffic arrival rate of femtocell users and macrocell users in a femtocell, respectively \\
${\lambda _T}$ & The total traffic arrival rate in a macrocell and its underlying ${N}$ femtocells \\
${\lambda _F}$ & The new traffic arrival rate of a femtocell user \\
${\lambda _{FU\_F}}$, ${\lambda _{FU\_M}}$ & The new traffic arrival rate of femtocell users originated from a femtocell and a macrocell, respectively \\
${\lambda_{FU\_FM}}$ & The new arrival rate of femtocell users in macrocell which hand off from femtocells  \\
${\lambda_{FU\_H}}$, ${\lambda_{MU\_H}}$ & The total handoff-in traffic arrival rate of femtocell users and macrocell users in a femtocell, respectively \\
${\lambda_{FU\_H}}$ & The total handoff traffic arrival rate from a macrocell into a femtocell by all active femtocell users \\
${\lambda_{FU\_MM}}$ & The handoff traffic arrival rate of all active femtocell users from an adjacent macrocell into the macrocell \\
${\lambda _M}$ & The total traffic arrival rate of all macrocell users \\
${\lambda _{MU\_F}}$, ${\lambda _{MU\_M}}$ & The new traffic arrival rate of macrocell users originated from a femtocell and a macrocell, respectively \\
${\lambda_{MU\_FM}}$ & The new arrival rate of macrocell users in macrocell which hand off from femtocells \\
${\lambda_{MU\_H}}$ & The total handoff traffic arrival rate from a femtocell into a macrocell by all active macrocell users \\
${\lambda_{MU\_MM}}$ & The handoff traffic arrival rate of all active macrocell users from an adjacent macrocell into the macrocell \\
${\mu _1}$, ${\mu _2}$ & The rate of channel holding time of femtocell users and macrocell users in a femtocell, respectively \\
$\sigma $ & The deviation parameter of log-normal shadowing \\
${1 \mathord{\left/ {\vphantom {1 {{\eta _{RT\_F}}}}} \right.\kern-\nulldelimiterspace} {{\eta _{RT\_F}}}}$, ${1 \mathord{\left/{\vphantom {1 {{\eta _{RT\_M}}}}} \right. \kern-\nulldelimiterspace} {{\eta_{RT\_M}}}}$ & The average dwelling time of femtoell users and macrocell users in a femtocell, respectively \\
${1 \mathord{\left/{\vphantom {1 \mu }} \right. \kern-\nulldelimiterspace} \mu }$ & The mean of user session duration \\
\hline
\end{tabular}
\end{table*}

\section{Markov Chain Model of Femtocell Networks}
\label{sec3}

In this section, the number of occupied femtocell channels in a femtocell is first modeled by a Markov chain. By analyzing the Markov chain model, the blocking probabilities for a femtocell user and a macrocell user are derived, respectively. Furthermore, the stationary state probabilities of the Markov chain model are derived to analyze the spectrum and energy efficiency of femtocell networks in the next section.

\subsection{Markov Chain State Transition Model}

To simplify the analysis using the Markov chain model, femtocells in a macrocell are considered as a homogeneous system where all femtocells have the same equipment parameters. Furthermore, this homogeneous system is assumed to be in a statistical equilibrium, which means the average hand-off arrival rate to a femtocell is equal to the corresponding hand-off departure rate. This allows a decoupling of a femtocell from its neighbors and permits an approximate analysis by consideration the femtocell and its overlaying macrocell \cite{Hu95}.

Let $\left( {i,j} \right)$
denote the two-dimension (2-D) state of Markov chain modeling femtocell channels usage within a femtocell, where $i$ represents
the number of femtocell channels including both open femotocell channels and closed femtocell channels, used by femtocell users and $j$ is the number of femtocell channels used by macrocell users in a femtocell. Fig. 2 illustrates the transition diagram
of a femtocell. Now, we describe the state transitions in detail.

\subsubsection{For $0 \le i \le {N_F} - {N_{F\_O}}$ and $0 \le j \le
{N_{F\_O}}$, the following transitions may occur}

\begin{itemize}
    \item $\left( {i,j} \right) \to \left( {i + 1,j} \right)$, when a closed
femtocell channel is occupied by a new femtocell user call originating from a femtocell or hand-off arrival to a femtocell, where $0 \le i < {N_F} - {N_{F\_O}}$.
    \item $\left( {i,j} \right) \to \left( {i,j + 1} \right)$, when an open
femtocell channel is occupied by a new macrocell user call originating from a femtocell or hand-off arrival to a femtocell, where $0 \le j <
{N_{F\_O}}$.
    \item $\left( {i,j} \right) \to \left( {i - 1,j} \right)$, when a closed
femtocell channel is released by a femtocell user call originating from a femtocell or hand-off departure from a femtocell, where $0 < i \le {N_F} - {N_{F\_O}}$.
    \item $\left( {i,j} \right) \to \left( {i,j - 1} \right)$, when an
open femtocell channel is released by a macrocell user call originating from a femtocell or hand-off departure from a femtocell, where $0 < j \le
{N_{F\_O}}$.
\end{itemize}
\subsubsection{For ${N_F} - {N_{F\_O}} \le i \le {N_F}$  and  $0 \le j \le {N_F} - i$,
the following transitions may occur}

\begin{itemize}
    \item $\left( {i,j} \right) \to \left( {i + 1,j} \right)$, when an open
femtocell channel is occupied by a new femtocell user call originating from a femtocell or hand-off arrival to a femtocell. This event occurs
when all closed femtocell channels are busy, where ${N_F} - {N_{F\_O}} \le i < {N_F}$.
    \item $\left( {i,j} \right) \to \left( {i,j + 1} \right)$, when an open
femtocell channel is occupied by a new macrocell user call originating from a femtocell or hand-off arrival to a femtocell, where $0 \le j < {N_F} - i$.
    \item $\left( {i,j} \right) \to \left( {i - 1,j} \right)$, when an open
femtocell channel is released by a femtocell user call originating from a femtocell or hand-off departure from a femtocell. This event occurs when the
femtocell user call has occupied an open femtocell channel, where ${N_F} - {N_{F\_O}} < i \le {N_F}$.
    \item $\left( {i,j} \right) \to \left( {i,j - 1} \right)$, when an open
femtocell channel is released by a macrocell user call originating from a femtocell or hand-off departure from a femtocell, where $0 < j \le {N_F} - i$.
\end{itemize}
\begin{figure}
\vspace{0.1in}
\centerline{\includegraphics[width=8cm,draft=false]{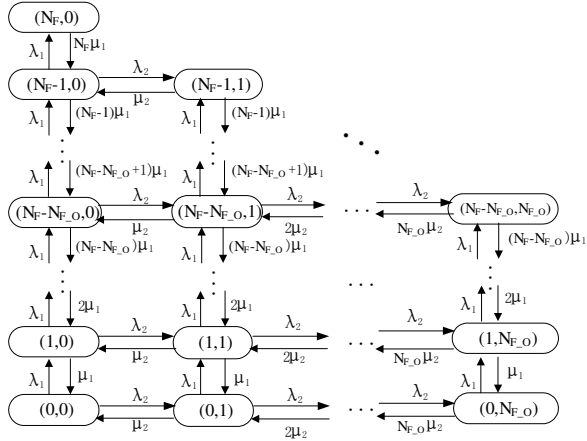}}
\caption{\small State transition diagram in a femtocell.}
\end{figure}

\subsection{Analysis of the Markov Chain Model}

In Fig. 1, we term a macrocell and its underlying $N$ femtocells as an entity
and assume that the total originating traffic process in an entity follows a
Poisson process with an arrive rate ${\lambda _T}$. For a femtocell user, it
may generate both indoor, i.e., femtocell, and outdoor, i.e., macrocell, call activities though indoor call activities occur more frequently. The fraction of indoor calls to the total number of calls originating from a femtocell user is denoted as $q$ and the new traffic arrival rate of a femtocell
user is denoted by ${\lambda _F}$. Hence, the total traffic arrival rate of all macrocell users in an entity is
${\lambda _M} = {\lambda _T} - NM{\lambda _F}$. The total traffic arrival rate of all macrocell users in a femtocell is $\frac{{{A_F}}}{{{A_M}}} \cdot {\lambda _M}$
and the total traffic arrival rate of all macrocell users in a macrocell is $\left( {1
- N\frac{{{A_F}}}{{{A_M}}}} \right){\lambda _M}$. The user session duration, which refers to the time duration of a requested session connection, is
assumed to be an exponential distribution with mean ${1 \mathord{\left/
{\vphantom {1 \mu }} \right. \kern-\nulldelimiterspace} \mu }$. A user dwelling time refers to the time
duration a user stays in a cell. The femtocell users and macrocell users dwelling time in a femtocell are assumed to follow an exponential distribution with mean
${1 \mathord{\left/ {\vphantom {1 {{\eta _{RT\_F}}}}} \right.
\kern-\nulldelimiterspace} {{\eta _{RT\_F}}}}$ and ${1 \mathord{\left/
{\vphantom {1 {{\eta _{RT\_M}}}}} \right. \kern-\nulldelimiterspace} {{\eta
_{RT\_M}}}}$, respectively.

The aggregate traffic arrival rate of femtocell users ${\lambda _1}$ in a femtocell  is
divided into two parts. One part is the new traffic arrival rate of femtocell users ${\lambda _{FU\_F}}$ originating from a femtocell and the other part is the total handoff-in traffic arrival rate of femtocell users ${\lambda
_{FU\_H}}$ caused by handoff from a macrocell into a femtocell by active
femtocell users. Therefore, the aggregate traffic arrival rate of femtocell users ${\lambda
_1}$ is determined by
\begin{equation}
{\lambda _1} = {\lambda _{FU\_F}} + {\lambda _{FU\_H}}.
\label{eq1}
\tag{1}
\end{equation}

Moreover, the new traffic arrival rate of femtocell users ${\lambda _{FU\_F}}$ originating from a femtocell is given by
\begin{equation}
{\lambda _{FU\_F}} = Mq{\lambda _F}.
\label{eq2}
\tag{2}
\end{equation}

Let ${P_{U\_M}}$ and ${P_{FU\_F}}$ denote the user blocking probability in a
macrocell and the blocking probability of femtocell users in a femtocell,
respectively. Based on the results in (2) and Appendix A, the aggregate traffic arrival rate of femtocell users ${\lambda _1}$ is derived as follows
\begin{equation}
\begin{split}
{\lambda _1} = Mq{\lambda _F} + \frac{1}{N}\left( {{\lambda _{FU\_M}} + {\lambda _{FU\_FM}} + {\lambda _{FU\_MM}}}\right)\\
\cdot \left( {1 - {P_{U\_M}}} \right) \cdot {P_{Handoff\_MF}},
\end{split}
\label{eq3}
\tag{3a}
\end{equation}
with
\begin{equation}
{\lambda _{FU\_M}} = NM\left( {1 - q} \right) \cdot {\lambda _F},
\label{eq4}
\tag{3b}
\end{equation}
\begin{equation}
{\lambda _{FU\_FM}} = N{\lambda _1} \cdot \left( {1 - {P_{FU\_F}}} \right) \cdot {P_{Handoff\_FM}},
\label{eq5}
\tag{3c}
\end{equation}
\begin{equation}
\begin{split}
{\lambda _{FU\_MM}} = \left( {{\lambda _{FU\_M}} + {\lambda _{FU\_FM}} + {\lambda _{FU\_MM}}} \right) \\
\cdot \left( {1 - {P_{U\_M}}} \right) \cdot {P_{Handoff\_MM}}.
\end{split}
\label{eq6}
\tag{3d}
\end{equation}

The channel holding time of femtocell users ${T_{cF\_F}}$ with rate ${\mu _1}$ in a femtocell is the minimum of the session duration and the average femtocell users dwelling time, i.e., ${T_{cF\_F}} = min({1 \mathord{\left/{\vphantom {1 \mu }} \right. \kern-\nulldelimiterspace} \mu },{1 \mathord{\left/ {\vphantom {1 {{\eta _{RT\_F}}}}} \right.\kern-\nulldelimiterspace} {{\eta _{RT\_F}}}})$. Then ${\mu _1}$ is given by \cite{Shun05}
\begin{equation}
{\mu _1} = \mu  + {\eta _{RT\_F}}.
\label{eq7}
\tag{4}
\end{equation}

The aggregate traffic arrival rate of macrocell users ${\lambda _2}$ in a femtocell is also divided into two parts. One part is the new traffic arrival rate of macrocell users ${\lambda _{MU\_F}}$ originating from femtocell itself and the other part is the handoff-in traffic arrival rate of macrocell users ${\lambda
_{MU\_H}}$ caused by handoff from a macrocell into a femtocell by active
macrocell users. Therefore, the aggregate traffic arrival rate of macrocell users ${\lambda _2}$ in a
femtocell is calculated by
\begin{equation}
{\lambda _2} = {\lambda _{MU\_F}} + {\lambda _{MU\_H}}.
\label{eq8}
\tag{5}
\end{equation}

Moreover, the new traffic arrival rate of macrocell users ${\lambda _{MU\_F}}$ is given by
\begin{equation}
{\lambda _{MU\_F}} = \frac{{{A_F}}}{{{A_M}}}{\lambda _M}.
\label{eq9}
\tag{6}
\end{equation}

Let ${P_{MU\_F}}$ denote the blocking probability of macrocell user in a
femtocell. Based on the results in (6) and Appendix B, the aggregate traffic arrival
rate of macrocell users ${\lambda _2}$ is derived as follows
\begin{equation}
\begin{split}
{\lambda _2} = \frac{{{A_F}}}{{{A_M}}}{\lambda _M} + \frac{1}{N}\left( {{\lambda _{MU\_M}} + {\lambda _{MU\_FM}} + {\lambda _{MU\_MM}}} \right) \\
\cdot \left( {1 - {P_{U\_M}}} \right) \cdot {P_{Handoff\_MF}},
\end{split}
\label{eq10}
\tag{7a}
\end{equation}
with
\begin{equation}
{\lambda _{MU\_M}} = \left( {1 - N\frac{{{A_F}}}{{{A_M}}}} \right) \cdot {\lambda _M},
\label{eq11}
\tag{7b}
\end{equation}
\begin{equation}
{\lambda _{MU\_FM}} = N{\lambda _2} \cdot \left( {1 - {P_{MU\_F}}} \right) \cdot {P_{Handoff\_FM}},
\label{eq12}
\tag{7c}
\end{equation}
\begin{equation}
\begin{split}
{\lambda _{MU\_MM}} = \left( {{\lambda _{MU\_M}} + {\lambda _{MU\_FM}} + {\lambda _{MU\_MM}}} \right) \\
\cdot \left( {1 - {P_{U\_M}}} \right) \cdot {P_{Handoff\_MM}}.
\end{split}
\label{eq13}
\tag{7d}
\end{equation}

The channel holding time of macrocell users ${T_{cM\_F}}$ with rate ${\mu _2}$ in a femtocell is the minimum of the session duration and the average macrocell users dwelling time, i.e., ${T_{cM\_F}} = min({1 \mathord{\left/{\vphantom {1 \mu }} \right. \kern-\nulldelimiterspace} \mu },{1 \mathord{\left/ {\vphantom {1 {{\eta _{RT\_F}}}}} \right.\kern-\nulldelimiterspace} {{\eta _{RT\_M}}}})$. Therefore, ${\mu _2}$ is given by \cite{Shun05}
\begin{equation}
{\mu _2} = \mu  + {\eta _{RT\_M}}.
\label{eq14}
\tag{8}
\end{equation}

\begin{figure*}[!t]
\[\left\{ \begin{array}{l}\left( {{\lambda _1} + {\lambda _2} + {\mu _1}i + {\mu _2}j} \right)S\left( {i,j} \right) = \left( {i + 1} \right){\mu _1}S\left( {i + 1,j} \right)\\{\rm{                                                                                            }} + \left( {j + 1} \right){\mu _2}S\left( {i,j + 1} \right) + {\lambda _1}S\left( {i - 1,j} \right) + {\lambda _2}S\left( {i,j - 1} \right),{\rm{     }}0 \le i \le {N_F},0 \le j \le \min \left( {{N_{F\_O}},{N_F} - i} \right)\\\left( {{\lambda _1} + {\lambda _2} + {\mu _1}i + {\mu _2}j} \right)S\left( {i,j} \right) = \left( {i + 1} \right){\mu _1}S\left( {i + 1,j} \right) + \left( {j + 1} \right){\mu _2}S\left( {i,j + 1} \right)\\ {\rm{                                          }} + {\lambda _2}S\left( {i,j - 1} \right),{\rm{                                           }}i = 0{\rm{ }}\\\left( {{\lambda _1} + {\lambda _2} + {\mu _1}i + {\mu _2}j} \right)S\left( {i,j} \right) = \left( {i + 1} \right){\mu _1}S\left( {i + 1,j} \right) + \left( {j + 1} \right){\mu _2}S\left( {i,j + 1} \right)\\{\rm{                                          }} + {\lambda _1}S\left( {i - 1,j} \right),{\rm{                                            }}j = 0\\\left( {{\lambda _1} + {\mu _1}i + {\mu _2}{N_{F\_O}}} \right)S\left( {i,j} \right) = \left( {i + 1} \right){\mu _1}S\left( {i + 1,j} \right) + {\lambda _1}S\left( {i - 1,j} \right)\\{\rm{                                          }} + {\lambda _2}S\left( {i,j - 1} \right),{\rm{                                            }}j = {N_{F\_O}}{\rm{            }}\\\left( {{\mu _1}i + {\mu _2}({N_F} - i)} \right)S\left( {i,j} \right) =  + {\lambda _1}S\left( {i - 1,j} \right) + {\lambda _2}S\left( {i,j - 1} \right),{\rm{                      }}i + j = {N_F}\end{array} \right. ,\tag{9}\]
\begin{equation}
\begin{split}
E\left( {T_{cM\_M}} \right) = \frac{1}{{\mu + {\eta _{RT\_M}}}}\left[ {N\frac{{{A_F}}}{{{A_M}}} \cdot {P_{MU\_F}}+\left( {1 - N\frac{{{A_F}}}{{{A_M}}}} \right)} \right]+ N\frac{{{A_F}}}{{{A_M}}}\left( {1 - {P_{MU\_F}}} \right) \cdot E\left( Z \right),
\end{split}
\label{eq22}
\tag{16a}
\end{equation}
\begin{equation}
E\left( Z \right) = \frac{1}{\mu } - \frac{1}{\mu }\left[ {\frac{{\ln \left( {\frac{\mu }{{{\eta _{RT\_M}}}}} \right)}}{{\frac{\mu }{{{\eta _{RT\_M}}}}}} - \frac{{{\eta _{RT\_M}}}}{\mu }\left( {{e^{ - \frac{{{\eta _{RT\_M}}}}{\mu }}} - 1} \right)} \right] ,
\label{eq23}
\tag{16b}
\end{equation}
\begin{equation}
\begin{split}
E\left( {T_{cF\_M}} \right) = \frac{1}{{\mu  + {\eta _{RT\_M}}}}\left[ {N\frac{{{A_F}}}{{{A_M}}} \cdot {P_{FU\_F}} + \left( {1 - N\frac{{{A_F}}}{{{A_M}}}} \right)} \right]+N\frac{{{A_F}}}{{{A_M}}}\left( {1 - {P_{FU\_F}}} \right) \cdot E\left( Z \right).
\end{split}
\label{eq25}
\tag{18}
\end{equation}
\end{figure*}

Based on the Markov chain state transition diagram in Fig. 2, the set of equilibrium
equations is given by (9), where $S\left( {i,j} \right)$ is the stationary state probability of Markov chain modeling femtocell channels in a femtocell. After
solving (9), a closed-form expression of the state probability $S\left( {i,j}
\right)$ is given by
\begin{equation}
S\left( {i,j} \right) = \frac{{{{\left( {{{{\lambda _1}} \mathord{\left/ {\vphantom {{{\lambda _1}} {{\mu _1}}}} \right. \kern-\nulldelimiterspace} {{\mu _1}}}} \right)}^i}{{\left( {{{{\lambda _2}} \mathord{\left/ {\vphantom {{{\lambda _2}} {{\mu _2}}}} \right. \kern-\nulldelimiterspace} {{\mu _2}}}} \right)}^j}}}{{i!j!}}S\left( {0,0} \right).
\label{eq16}
\tag{10}
\end{equation}

The normalization condition is given by
\begin{equation}
\sum\limits_{\left( {i,j} \right)} {S\left( {i,j} \right) = 1}.
\label{eq17}
\tag{11}
\end{equation}

From (11), the idle-state probability $S\left( {0,0} \right)$, i.e. the probability that there is no active user using the channel, is derived as follows
\begin{equation}
S\left( {0,0} \right) = {\left\{ {\sum\limits_{i = 0}^{{N_F}} {\sum\limits_{j = 0}^{j = \min \left( {{N_{F\_O}},{N_F} - i} \right)} {\frac{{{{\left( {{{{\lambda _1}} \mathord{\left/ {\vphantom {{{\lambda _1}} {{\mu _1}}}} \right. \kern-\nulldelimiterspace} {{\mu _1}}}} \right)}^i}{{\left( {{{{\lambda _2}} \mathord{\left/ {\vphantom {{{\lambda _2}} {{\mu _2}}}} \right. \kern-\nulldelimiterspace} {{\mu _2}}}} \right)}^j}}}{{i!j!}}} } } \right\}^{ - 1}}.
\label{eq18}
\tag{12}
\end{equation}

As a consequence of the above equation, all other state probabilities can be obtained by
substituting (12) into (10).

Note that the focus of this paper is on studying the call blocking probability where an incoming call is blocked when there is no channel available in the network to serve the call. Based on the Markov chain state transition diagram in Fig. 2 and (10), the blocking probability of femtocell user in a femtocell is given by
\begin{equation}
{P_{FU\_F}} = \sum\limits_{i = {N_F} - {N_{F\_O}}}^{i = {N_F}} {S\left( {i,{N_F} - i} \right)} .
\label{eq19}
\tag{13}
\end{equation}

Moreover, the blocking probability of macrocell user in a femtocell is given by
\begin{equation}
\begin{split}
{P_{MU\_F}} = \sum\limits_{i = 0}^{i = {N_F} - {N_{F\_O}} - 1} {S\left( {i,{N_{F\_O}}} \right)}\\
+ \sum\limits_{i = {N_F} - {N_{F\_O}}}^{i = {N_F}} {S\left( {i,{N_F} - i} \right)} .
\end{split}
\label{eq20}
\tag{14}
\end{equation}

Based on the Erlang-B formula in \cite{Kleinrock75}, the user blocking probability in a macrocell is given by
\begin{equation}
{P_{U\_M}} = \frac{{{{\left( {{T_M} + {T_F}} \right)}^{{N_M}}}/{N_M}!}}{{\sum\nolimits_{k = 0}^{{N_M}} {{{\left( {{T_M} + {T_F}} \right)}^k}/k!} }},
\label{eq21}
\tag{15}
\end{equation}
where ${T_M}$ and ${T_F}$ are traffic intensities of macrocell users and
femtocell users in a macrocell, respectively. ${N_M}$ is the number of
channels in a macrocell. Let ${T_{cM\_M}}$ denotes the channel holding time of a macrocell
user in a macrocell. It can be readily shown that the expectation of ${T_{cM\_M}}$ is given by (16) \cite{Zhang10},
where ${E\left[\cdot\right]}$ denotes an expectation operator. Then, the traffic intensity of macrocell users ${T_M}$ in a macrocell
is calculated by
\begin{equation}
{T_M} = \left( {{\lambda _{MU\_M}} + {\lambda _{MU\_FM}} + {\lambda _{MU\_MM}}} \right) \cdot E\left( {T_{cM\_M}} \right).
\label{eq24}
\tag{17}
\end{equation}

Let ${T_{cF\_M}}$ denotes the channel holding time of a femtocell user in a macrocell.
The expectation of ${T_{cF\_M}}$ is given by (18) \cite{Zhang10} .

Furthermore, the traffic intensity of femtocell user ${T_F}$ in a macrocell is given by
\begin{equation}
{T_F} = \left( {{\lambda _{FU\_M}} + {\lambda _{FU\_FM}} + {\lambda _{FU\_MM}}} \right) \cdot E\left( {T_{cF\_M}} \right).
\label{eq26}
\tag{19}
\end{equation}

Substituting (17) and (19) into (15), a closed-form user blocking probability in a
macrocell is derived. Equations (13), (14) and (15) represent a nonlinear
system, which can be solved by numerical techniques.

\subsection{Analysis of blocking probabilities}

Based on the proposed blocking probability models for two-tier femtocell
networks, performance evaluation is analyzed as follows. Unless otherwise specified, the key
parameters are configured as: ${R_M} = 1000m$, ${R_F} = 20m$, $N = 40$, ${N_F} = 3$, ${N_{F\_O}} = 1$, ${N_M} = 24$,  $q = 0.6$, ${\lambda _F} =
0.002$ and $M$ is 4, 6 and 8 \cite{Mansfield08,ABI07}. The average value of the user session duration is set
as ${1 \mathord{\left/ {\vphantom {1 \mu }} \right.
\kern-\nulldelimiterspace} \mu } = 110$ seconds \cite{3g}. The average values of femtocell users and macrocell users dwelling time in a femtocell are configured as ${1 \mathord{\left/
{\vphantom {1 {{\eta _{RT\_F}}}}} \right. \kern-\nulldelimiterspace} {{\eta
_{RT\_F}}}} = 990$ and ${1 \mathord{\left/ {\vphantom {1 {{\eta _{RT\_M}}}}}
\right. \kern-\nulldelimiterspace} {{\eta _{RT\_M}}}} = 300$ seconds \cite{Zhang10,Xiang10},
respectively. To validate the proposed model, we compare the proposed model with the model derived from the reference \cite{Zhang10} in Fig. 3 and Fig. 4.

Fig. 3 shows the user blocking probability ${P_{U\_M}}$ in a macrocell in terms of the total
arrival rate ${\lambda _T}$ with different average number of femtocell users in a
femtocell. The curves in Fig. 3 illustrate that the value of ${P_{U\_M}}$
increases with an increase in the arrival rate ${\lambda _T}$. When the number of
occupied channels increases as a consequence of an increase in the arrival rate in a macrocell, the user
blocking probability in a macrocell ${P_{U\_M}}$ increases. When the total arrival
rate ${\lambda _T}$ is fixed, it is interesting to see that the user blocking probability ${P_{U\_M}}$ decreases
with an increase in the average number of femtocell users in a femtocell. This is caused by that an increase in the average number of femtocell users in a femtocell will
reduce the traffic flow density in a macrocell when the total arrival rate ${\lambda
_T}$ is fixed. As a consequence, the user blocking
probability ${P_{U\_M}}$ decreases when more users are added into a
femtocell. The curves obtained from the proposed model exhibit a good match with the curves obtained from the model in reference \cite{Zhang10}. This result indicates that the proposed model is consistent with the model in reference [8] if the impact of closed and open femtocell channels is not considered in simulation results.

Fig. 4 illustrates the blocking probability of femtocell users  $P_{FU\_M}$ in terms of the total arrival rate ${\lambda _T}$ with different average number of femtocell users in a femtocell. The curves in Fig. 4 show that the blocking probability of femtocell users $P_{FU\_M}$ increases with the higher total arrival rate${\lambda _T}$. We can also observe that, when the total arrival rate ${\lambda _T}$ is fixed, the blocking probabilities of femtocell users increase with an increase in the average number of femtocell users. Compared with the curves obtained from the model in reference \cite{Zhang10}, the curves obtained from the proposed model demonstrate the same trend in Fig. 4. In the proposed model, femtocell channels are not only open for femtocell users but also partially open to macrocell users. In the model derived from \cite{Zhang10}, femtocell channels are only open for femtocell users. In this case, the blocking probability derived from the proposed model is higher than the blocking probability in \cite{Zhang10}.

Fig. 5 analyzes the blocking probability of macrocell users $P_{MU\_M}$ in terms of the total arrival rate $\lambda_T$ with different average number of femtocell users in a femtocell. Numerical results show that the blocking probability of macrocell users $P_{MU\_M}$ increases with the higher total arrival rate $\lambda_T$. Moreover, when the total arrival rate $\lambda_T$ is fixed, the blocking probability of macrocell users increases with an increase in the average number of femtocell users in a femtocell.

\begin{figure}
\vspace{0.1in}
\centerline{\includegraphics[width=8cm,draft=false]{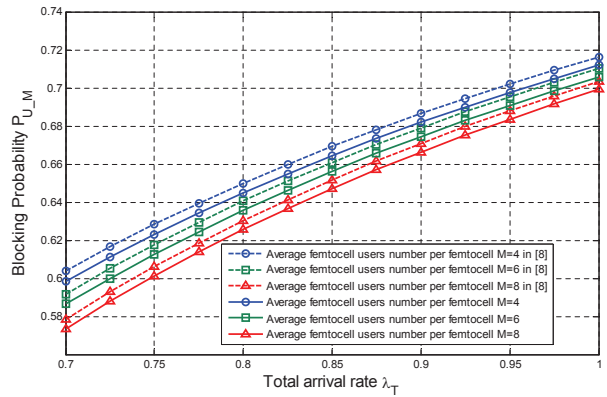}}
\caption{\small The user blocking probability in terms of
total arrival rate with different average number of femtocell users.}
\end{figure}
\begin{figure}
\vspace{0.1in}
\centerline{\includegraphics[width=8cm,draft=false]{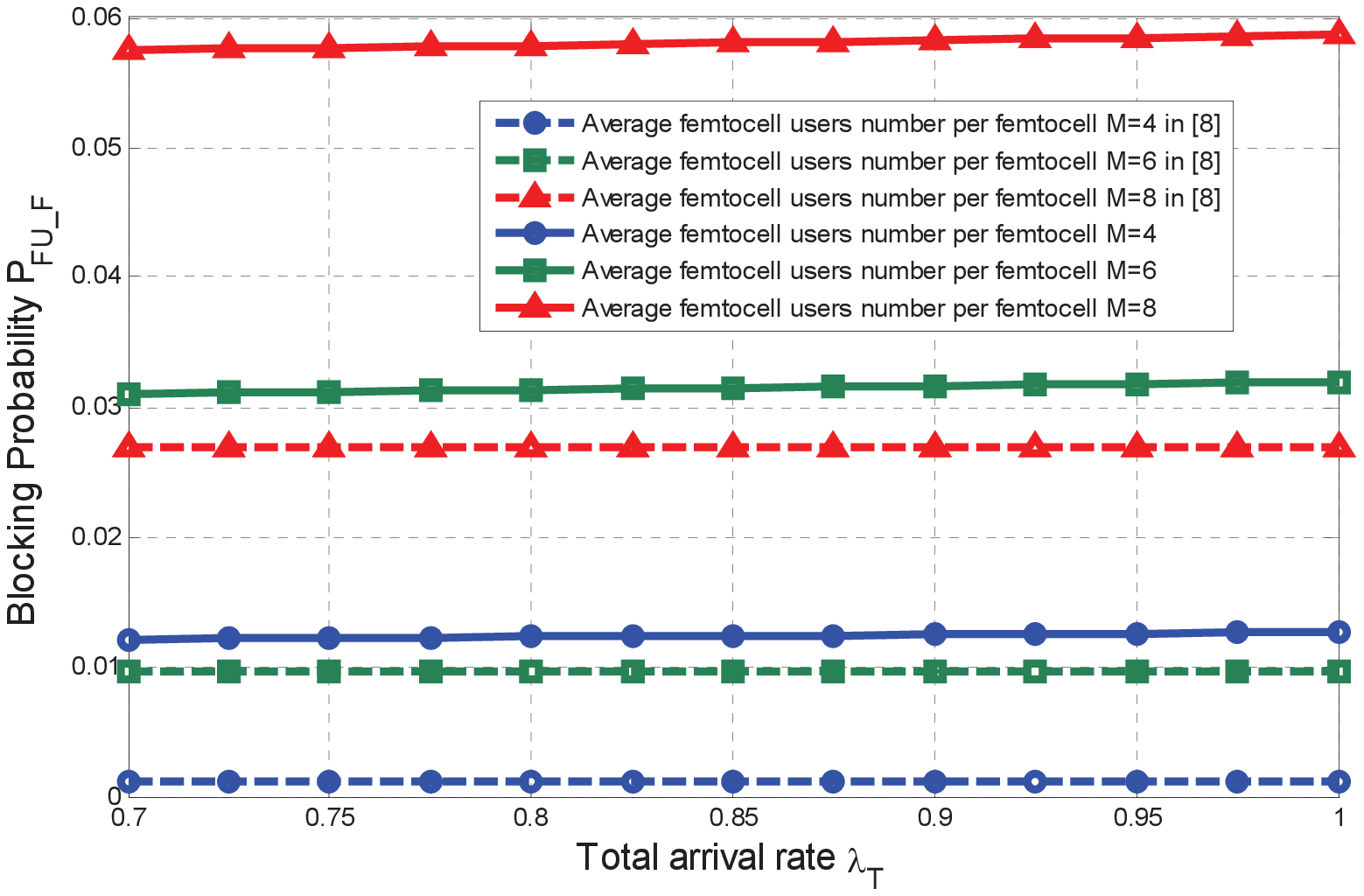}}
\caption{\small  The blocking probability of femtocell users in terms of
total arrival rate with different average number of femtocell users.}
\end{figure}
\begin{figure}
\vspace{0.1in}
\centerline{\includegraphics[width=8cm,draft=false]{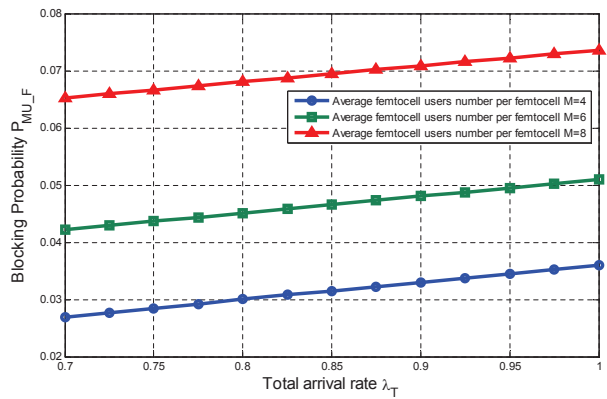}}
\caption{\small  The blocking probability of macrocell users in terms of
total arrival rate with different average number of femtocell users.}
\end{figure}

\section{Spectrum and Energy Efficiency Models of Femtocell Networks}
\label{sec4}

Based on the stationary state probabilities derived by the Markov chain models, the spectrum and energy efficiency models are proposed for femtocell networks. Furthermore, simulations are performed to analyze the performance parameters of spectrum and energy efficiency in femtocell networks.

\subsection{Spectrum Efficiency Model of Femtocell Networks}

The occupancy probability of a closed femtocell channel is analyzed by considering two situations separately. In the first situation, the number of channels occupied by femtocell users is smaller than or equal to the number of closed femtocell channels in a femtocell; in the second situation, the number of channels occupied by femtocell users is larger than the number of closed femtocell channels in a femtocell. Therefore, using (10), the occupancy probability of a closed femtocell channel is given by
\begin{equation}
\begin{split}
{P_{closed}} = \sum\limits_{i = 1}^{{N_F} - {N_{F\_O}}} {\sum\limits_{j = 0}^{{N_{F\_O}}} {\frac{{\binom{{N_F} - {N_{F\_O}} - 1}{i - 1}}}{{\binom{{N_F} - {N_{F\_O}}}{i}}}} } S\left( {i,j} \right)\\
 + \sum\limits_{i = {N_F} - {N_{F\_O}} + 1}^{{N_F}} {\sum\limits_{j = 0}^{{N_F} - i} {S\left( {i,j} \right)} } ,
\end{split}
\label{eq27}
\tag{20}
\end{equation}
where $\binom{x}{y}$ is the binomial coefficients with parameters $x$ and $y$. Similarly, the occupancy probability of an open femtocell channel can also be analyzed by considering two situations separately. The first situation occurs when the femtocell
users do not occupy the open femtocell channels in a femtocell. The second situation occurs when the femtocell users occupy the open femtocell channels in a femtocell.
Therefore, using (10), the occupancy probability of an open femtocell channel is given by
\begin{equation}
\begin{split}
{P_{open}} = \sum\limits_{i = 0}^{{N_F} - {N_{F\_O}}} {\sum\limits_{j = 1}^{{N_{F\_O}}} {\frac{{\binom{{N_{F\_O}} - 1}{j - 1}}}{{\binom{{N_{F\_O}}}{j}}}} } S\left( {i,j} \right) \\
+ \sum\limits_{i = {N_F} - {N_{F\_O}} + 1}^{{N_F}} {\sum\limits_{j = 0}^{{N_F} - i} {\frac{{\binom{{N_{F\_O}} - 1}{i - ({N_F} - {N_{F\_O}}) - 1}}}{{\binom{{N_{F\_O}}}{i - ({N_F} - {N_{F\_O}})}}}} } S\left( {i,j} \right) ,
\end{split}
\label{eq28}
\tag{21}
\end{equation}

Considering that femtocells are assumed to be uniformly distributed in a macrocell, the
probability density function (PDF) of the distance $l$ between two femtocell
BSs in a macrocell is given \cite{Bettstetter02}
\begin{equation}
  f(l) = \frac{{4l}}{{\pi {R_M}^2}}\left( {\arccos \left( {\frac{l}{{2{R_M}}}} \right) - \frac{l}{{2{R_M}}}\sqrt {1 - {{\left( {\frac{l}{{2{R_M}}}} \right)}^2}} } \right)
  \label{eq29}
  \tag{22}
\end{equation}
where $0 \le l \le 2{R_M}$. Moreover, considering that the distance between a user and its associated femtocell BS is usually small, the distance between a user and the interfering femtocell BS is approximated by the distance between the associated femtocell BS and the interfering femtocell BS.

To avoid co-channel interference between macrocells and femtocells, it is assumed that macrocells and femtocells use different frequencies for communications. Furthermore, all femtocells in a macrocell share the same frequency bandwidth and the frequency bandwidth used by a femtocell is divided into ${N_F}$
no-overlapping sub-bandwidth. Here, each frequency sub-bandwidth corresponds to a femtocell
channel. In this case, the number of interfering users from adjacent
femtocells is no more than the number of femtocells in a macrocell. For the sake of illustration,
the number of interfering users from an adjacent femtocell is configured as one. Since the
radius of a macrocell is much larger than the radius of a femtocell, the
co-channel interference from adjacent macrocells is ignored in this paper.
Furthermore, a user in a femtocell is only interfered by users using same
frequency sub-bandwidth in adjacent femtocells and located in the same macrocell. We consider
the propagation effects of path loss, shadowing, and Rayleigh fading over femtocell channels.
When an active user is located in a femtocell, the active user receives interference from the ${k^{th}}$ adjacent femtocell, $1 \le k \le {N-1}$.
The interference power originating from the ${k^{th}}$ adjacent femtocell is expressed as \cite{802.16m,Cheng12,Ge11}
\begin{equation}
  {I_k} = P{W_v}\frac{{{e^{2\sigma G}}{\alpha ^2}}}{{{{10}^{1.2{n_w}}}{l^\beta }}},
  \label{eq30}
  \tag{23}
\end{equation}
where $P{W_v}$ is the transmission power over a femtocell channel. The total
transmission power of a femtocell is distributed over all femtocell channels.
The term ${e^{2\sigma G}}$ accounts for log-normal shadowing with deviation
$\sigma $, where $G \sim Gaussian\left( {0,1} \right)$ represents a standard
normal random variable. The random variable ${\alpha ^2}$ is exponentially distributed
with mean value 1 in a Rayleigh fading environment. The item ${10^{1.2{n_w}}}$ refers
to the through-wall loss in an indoor environment with the number of walls
among femtocells ${n_w}$. The item ${l^\beta }$ stands for the path loss
effect between two femtocells within a macrocell with path loss exponent
$\beta$.

Considering that femtocells are usually used for indoor environment, the signal received by a user in a femtocell does not consider the small
scale fading and the through-wall loss. Therefore, the desired signal power ${S_m}$
received by the ${m^{th}}$ user in a femtocell is given by
\begin{equation}
  {S_m} = P{W_v}\frac{{{Z_{shadowing}}}}{{{L^\beta }}},
  \label{eq31}
  \tag{24}
\end{equation}
where ${Z_{shadowing}}$ indicates the shadowing effect in an indoor
environment and is assumed to be 4dB \cite{femtoforum}. $L$ denotes the distance between
the user $U{E_m}$ and the corresponding femtocell BS. Users are uniformly distributed in a
femtocell and the protection distance between the user and the corresponding femtocell BS
is ${R_p}$. The PDF of $L$ is given by
\begin{equation}
f\left( L \right) = \frac{{2L}}{{{R_F}^2}}
\label{eq32}
\tag{25}
\end{equation}
where ${R_p} \le L \le {R_F}$.

Furthermore, the capacity of all closed channels in a femtocell is derived by (26),
where ${n_0}$ denotes the additive white Gaussian noise (AWGN) in wireless
channels. ${B_W}$ represents the bandwidth of femtocell channel. The capacity
of all open channels in a femtocell is derived by (27).
As a consequence, the total capacity of a femtocell is derived by (28).
\begin{figure*}[!t]
\[{C_{closed}} = \sum\limits_{m = 1}^{{N_F} - {N_{F\_O}}} {{B_W}\log \left( {1 + \frac{{{P_{closed}}{S_m}}}{{{n_0} + \sum\limits_{n = 1}^{N - 1} {\binom{N - 1}{n}{{\left( {\frac{{{P_{closed}}}}{{{N_F} - {N_{F\_O}}}}} \right)}^n}{{\left( {1 - \frac{{{P_{closed}}}}{{{N_F} - {N_{F\_O}}}}} \right)}^{N - 1 - n}}\sum\limits_{k = 1}^n {{I_k}} } }}} \right)} ,\tag{26} \]
\[{C_{open}} = \sum\limits_{m = 1}^{{N_{F\_O}}} {{B_W}\log \left( {1 + \frac{{{P_{open}}{S_m}}}{{{n_0} + \sum\limits_{n = 1}^{N - 1} {\binom{N - 1}{n}{{\left( {\frac{{{P_{open}}}}{{{N_{F\_O}}}}} \right)}^n}{{\left( {1 - \frac{{{P_{open}}}}{{{N_{F\_O}}}}} \right)}^{N - 1 - n}}\sum\limits_{k = 1}^n {{I_k}} } }}} \right)}.\tag{27}\]
\[\begin{array}{l}{C_{total}} = {C_{closed}} + {C_{open}}= \sum\limits_{m = 1}^{{N_F} - {N_{F\_O}}} {{B_W}\log \left( {1 + \frac{{{P_{closed}}{S_m}}}{{{n_0} + \sum\limits_{n = 1}^{N - 1} {\binom{N - 1}{n}{{\left( {\frac{{{P_{closed}}}}{{{N_F} - {N_{F\_O}}}}} \right)}^n}{{\left( {1 - \frac{{{P_{closed}}}}{{{N_F} - {N_{F\_O}}}}} \right)}^{N - 1 - n}}\sum\limits_{k = 1}^n {{I_k}} } }}} \right)} \\{\rm{        }} + \sum\limits_{m = 1}^{{N_{F\_O}}} {{B_W}\log \left( {1 + \frac{{{P_{open}}{S_m}}}{{{n_0} + \sum\limits_{n = 1}^{N - 1} {\binom{N - 1}{n}{{\left( {\frac{{{P_{open}}}}{{{N_{F\_O}}}}} \right)}^n}{{\left( {1 - \frac{{{P_{open}}}}{{{N_{F\_O}}}}} \right)}^{N - 1 - n}}\sum\limits_{k = 1}^n {{I_k}} } }}} \right)} \end{array}. \tag{28}\]
\[{\eta _{EE}} = \frac{{P{W_c} + \sum\limits_{i = 0}^{{N_F}} {\sum\limits_{j = 0}^{\min \left( {{N_{F\_O}},{N_F} - i} \right)} {(i + j)S(i,j)} } P{W_v}}}{{{C_{total}}}},\tag{32a}\]
\[\begin{array}{l}{C_{total}} = \sum\limits_{m = 1}^{{N_F} - {N_{F\_O}}} {{B_W}\log \left( {1 + \frac{{{P_{closed}}{S_m}}}{{{n_0} + \sum\limits_{n = 1}^{N - 1} {\binom{N - 1}{n}{{\left( {\frac{{{P_{closed}}}}{{{N_F} - {N_{F\_O}}}}} \right)}^n}{{\left( {1 - \frac{{{P_{closed}}}}{{{N_F} - {N_{F\_O}}}}} \right)}^{N - 1 - n}}\sum\limits_{k = 1}^n {{I_k}} } }}} \right)} \\{\rm{               }} + \sum\limits_{m = 1}^{{N_{F\_O}}} {{B_W}\log \left( {1 + \frac{{{P_{open}}{S_m}}}{{{n_0} + \sum\limits_{n = 1}^{N - 1} {\binom{N - 1}{n}{{\left( {\frac{{{P_{open}}}}{{{N_{F\_O}}}}} \right)}^n}{{\left( {1 - \frac{{{P_{open}}}}{{{N_{F\_O}}}}} \right)}^{N - 1 - n}}\sum\limits_{k = 1}^n {{I_k}} } }}} \right)} \end{array}. \tag{32b}\]
\end{figure*}

\subsection{Energy Efficiency Model of Femtocell Networks}

A femtocell BS energy consumption can be decomposed into the fixed energy
consumption part and the dynamic energy consumption part \cite{Humar11}. The fixed
energy consumption, e.g., the circuit energy consumption, is the baseline
energy consumed at a femtocell BS. The circuit energy consumption usually
depends on both hardware and software configurations of a femtocell BS and is
independent of the number of occupied channels. The dynamic energy
consumption accounts for the transmission energy consumed in radio
frequency (RF) transmission circuits depending on the number of occupied
channels. As an easy consequence of the above analysis, we can
build a femtocell BS energy consumption model
\[E\left( {P{W_{FBS}}} \right) = P{W_c} + E\left( {P{W_t}} \right) ,\tag{29}\]
where $P{W_{FBS}}$ denotes the total energy consumption of femtocell BS.
$P{W_{c}}$ refers to the fixed energy consumption of femtocell BS. $P{W_{t}}$
indicates the dynamic energy consumption of femtocell BS. The dynamic energy
consumption is mainly associated with the transmission energy over wireless
channels. Based on the Markov chain state transition diagram in Fig. 2, the
average dynamic energy consumption is derived as follows
\[E\left( {P{W_t}} \right) = \sum\limits_{i = 0}^{{N_F}} {\sum\limits_{j = 0}^{\min \left( {{N_{F\_O}},{N_F} - i} \right)} {(i + j)S(i,j)} } P{W_v} .\tag{30}\]

It is very important to study the spectrum and energy efficiency from a systematic
perspective. For this, we introduce a new performance metric \textit{the
utility function of energy efficiency}, which is defined as the ratio of the
total capacity in a femtocell to the average total energy consumption in a
femtocell BS. Let ${\eta _{EE}}$ denote the utility function of energy
efficiency. Then, we have
\[{\eta _{EE}} = \frac{{E\left( {P{W_{FBS}}} \right)}}{{{C_{total}}}} .\tag{31}\]

Based on (28) and (29), the energy efficiency model can be further derived as
follows (32).

\section{Numerical Results and Discussions}
\label{sec5}

In this subsections numerical and MC simulations are presented to demonstrate interactions
between the femtocell energy efficiency metrics and critical performance-impacting parameters. Unless otherwise specified, the following parameters are used in the numerical and MC simulations: $\sigma  = 8dB$,
$\beta  = 2$,  ${n_w} = 2$, and  ${R_p} = 5m$.

First, we fix the number of total femtocell channels ${N_F=6}$ and the number of open femtocell channels ${N_{F\_O}=3}$. Fig. 6 illustrates the spectrum efficiency of the femtocell networks in terms of the average number of femtocells users with different number of femtocells in a
macrocell, in which "Num" labels the numerical results and "MC" represents the MC simulation results. When the number of femtocells in a macrocell is fixed, the
spectrum efficiency increases with an increase in the average number of users in a femtocell. When the average number
of users in a femtocell is fixed, the spectrum efficiency of femtocell
networks decreases with an increase in the number of femtocells in a macrocell. Fig. 7 shows the
energy efficiency of the femtocell networks in terms of the average number of
femtocell users with different number of femtocells in a macrocell. When the
number of femtocells in a macrocell is fixed, the energy efficiency of
femtocell networks increases with an increase in the average number of users in a femtocell. When the average number
of users in a femtocell is fixed, the energy efficiency of femtocell networks
increases with an increase in the number of femtocells in a macrocell. Compared with results from MC simulations, these numerical results are validated in Fig. 6 and Fig. 7, which demonstrate good accuracy of the results.

Secondly, we fix the number of open femtocell channels ${N_{F\_O}=3}$ and the average number of users as 4 in a femtocell. Fig. 8 shows the spectrum efficiency in terms of the number of closed channels in a femtocell with different number of femtocells in a macrocell.
It is observed that the spectrum efficiency increases with an increase in the number of closed
channels in a femtocell. In addition, when the number of closed channels is fixed, the
spectrum efficiency of the femtocell networks decreases with an increase in the number of femtocells
in a macrocell. Fig. 9 shows the energy efficiency performance in terms of
the number of closed channels in a femtocell. We can observe that the energy
efficiency of the femtocell networks decreases with an increase in the number of closed channels in a
femtocell. On the other hand, the energy efficiency increases with an increase in the number of
femtocells in a macrocell. The numerical results are validated by the MC simulation results shown in Fig. 8 and Fig. 9. However, the MC simulation curves are less than the numerical curves in Fig. 8 and Fig. 9. Considering that the MC simulation results are realized by finite simulation calculations, a few MC simulation calculation values with small probabilities will be discarded in the final results.

In the end, we fix the number of closed femtocell channels to be 2 and the number of femtocells in a macrocell to be 25 for the following simulation. Fig. 10 shows the spectrum efficiency of the femtocell networks in terms of
the average number of femtocell users with different number of open channels in a femtocell. In
this example, the number of total channels in a femtocell is set as 8. We can
see that the spectrum efficiency of the femtocell networks increases with
an increase in the average number of users in a femtocell. When the average number of users is fixed in a femtocell, the spectrum efficiency increases with an increase in the number of open
channels in a femtocell. Fig. 11 shows the energy efficiency of the femtocell
networks in terms of the average number of femtocell users with different open
channels in a femtocell. The curves show that the energy efficiency
increases with an increase in the average number of users in a femtocell and the energy efficiency decreases with an increase in the number of open channels in a femtocell.
According to the energy efficiency model in (32), the average dynamic energy consumption linearly increases with an increase in the number of open femtocell channels but the total capacity of
a femtocell only increases logarithmically with an increase in the number of open femtocell channels. Therefore, when the number of open femtocell channels increases, the energy efficiency decreases. Based on results of Fig. 8-11, our analytical models and simulation results indicate that an increase in the number of open or closed femtocell channels can
conduce to the increase of spectrum efficiency and the decrease of energy efficiency in a two-tier femtocell networks. As a consequence, the results provide guidelines for trading off the spectrum
and energy efficiency of two-tier femtocell networks by configuring different number of open or closed femtocell channels in a femtocell.
\begin{figure}
\vspace{0.1in}
\centerline{\includegraphics[width=8cm,draft=false]{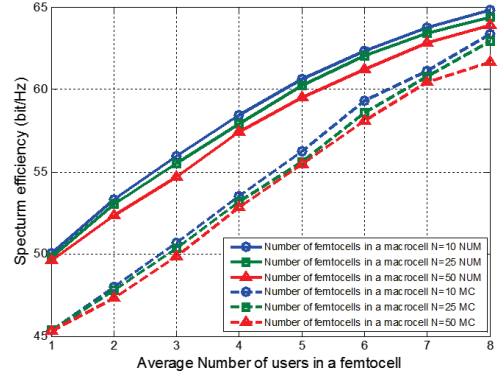}}
\caption{\small  Spectrum efficiency of femtocell networks with respect to
the average number of users in a femtocell.}
\end{figure}
\begin{figure}
\vspace{0.1in}
\centerline{\includegraphics[width=8cm,draft=false]{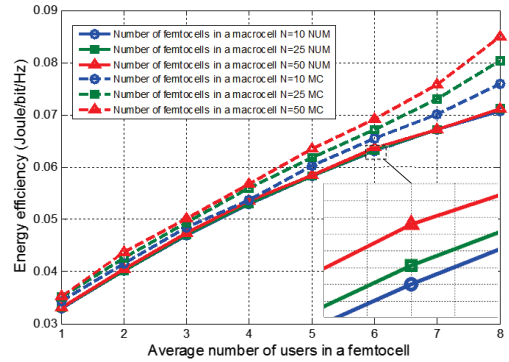}}
\caption{\small  Energy efficiency of femtocell networks with respect to the
average number of users in a femtocell.}
\end{figure}
\begin{figure}
\vspace{0.1in}
\centerline{\includegraphics[width=8cm,draft=false]{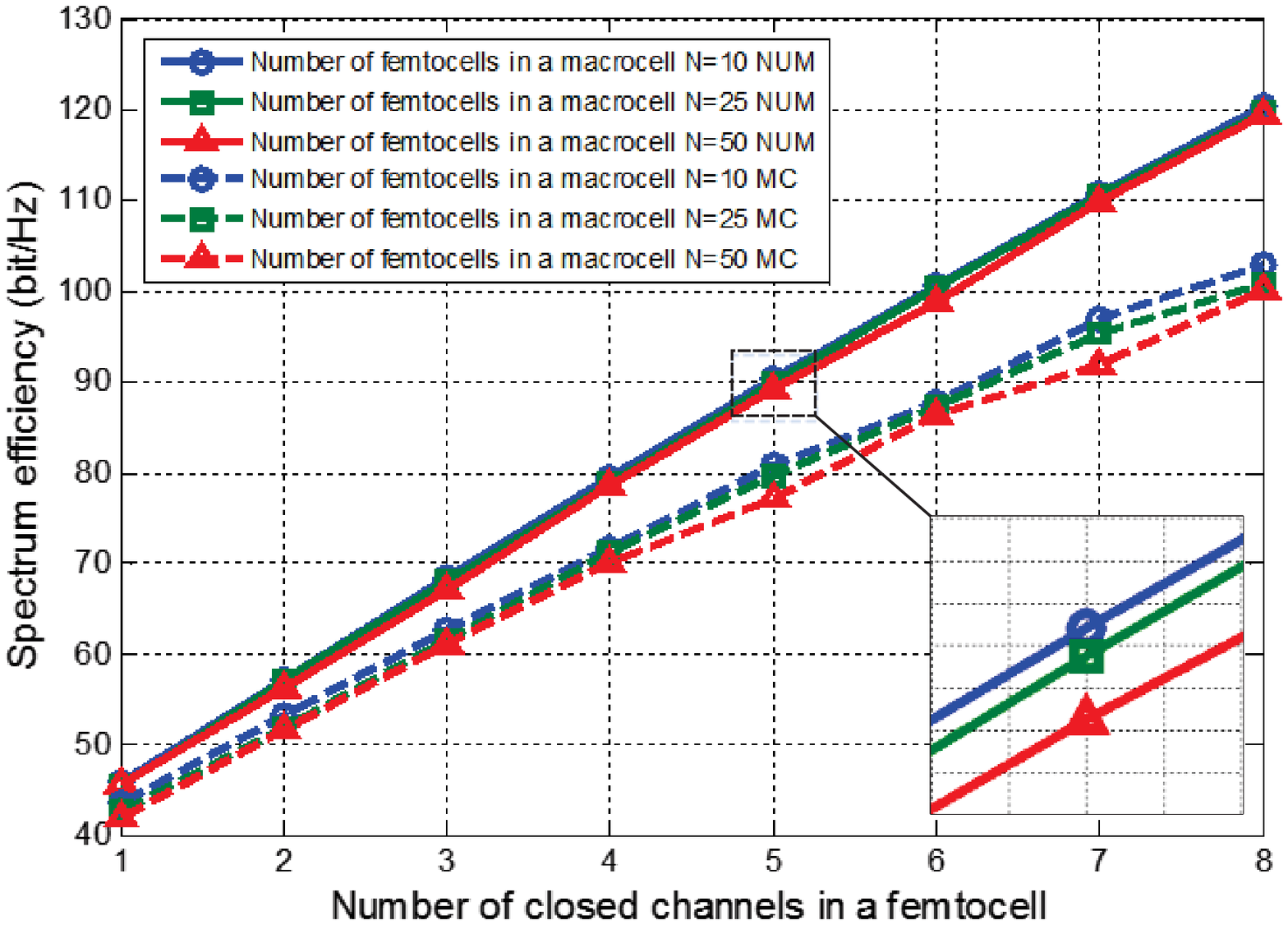}}
\caption{\small  Spectrum efficiency of femtocell networks with respect to
the number of closed channels in a femtocell.}
\end{figure}
\begin{figure}
\vspace{0.1in}
\centerline{\includegraphics[width=8cm,draft=false]{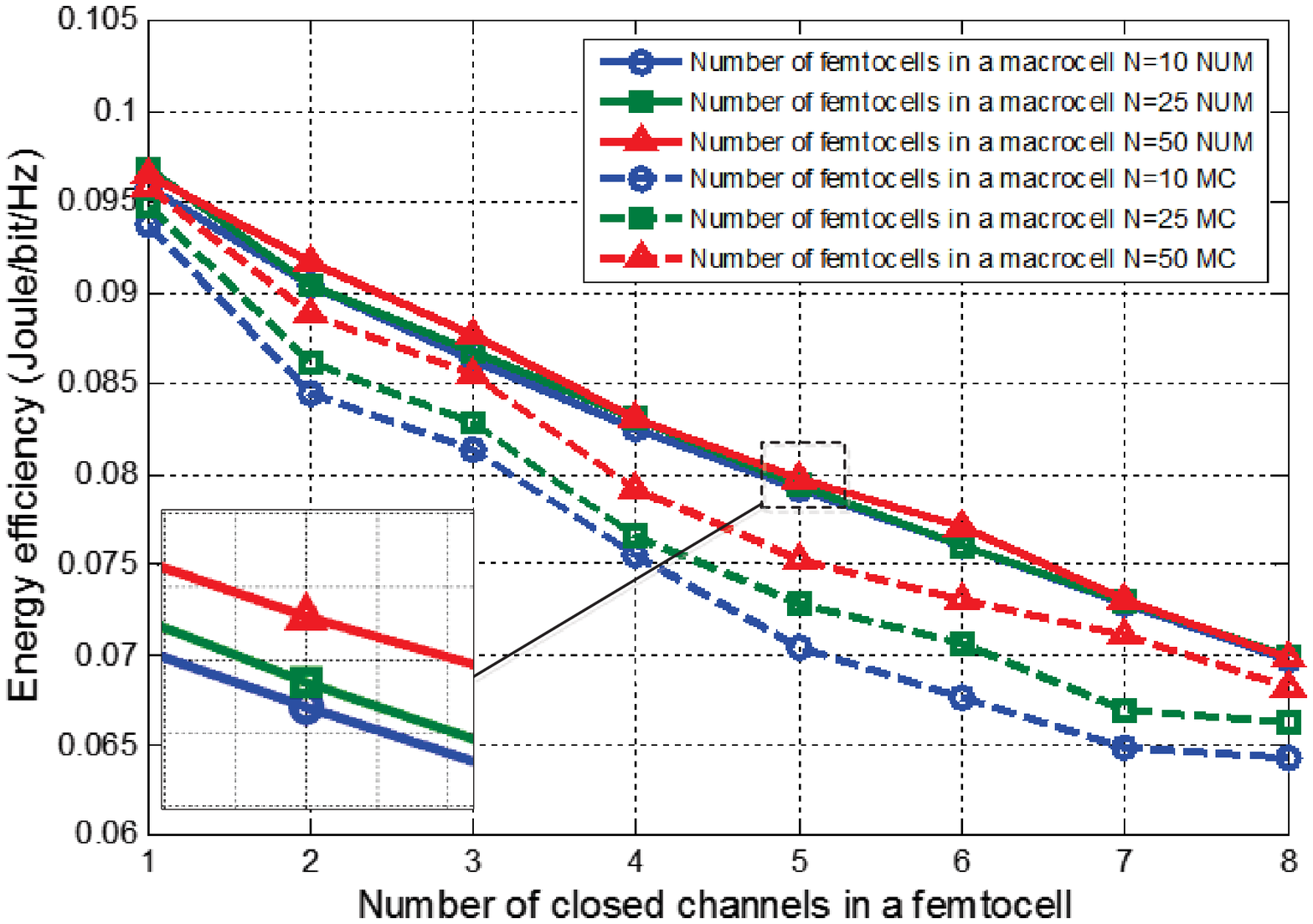}}
\caption{\small  Energy efficiency of femtocell networks with respect to the
number of closed channels in a femtocell.}
\end{figure}
\begin{figure}
\vspace{0.1in}
\centerline{\includegraphics[width=8cm,draft=false]{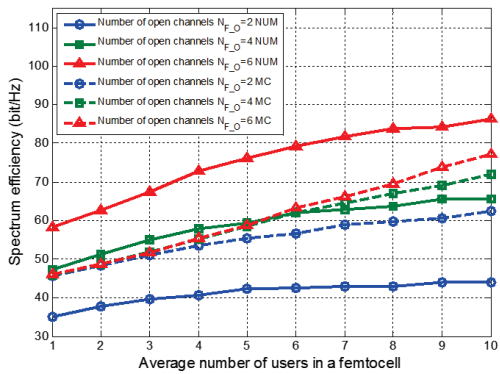}}
\caption{\small Spectrum efficiency of femtocell networks with respect to the
number of open channels.}
\end{figure}
\begin{figure}
\vspace{0.1in}
\centerline{\includegraphics[width=8cm,draft=false]{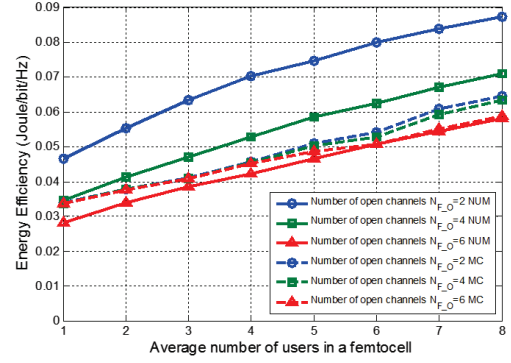}}
\caption{\small  Energy efficiency of femtocell networks with respect to the
number of open channels.}
\end{figure}

\section{Conclusions}
\label{sec6}

In this paper, the channel occupancy in a femtocell is modeled by Markov chains. To derive state
transition probabilities in a femtocell, a Markov chain state transition
diagram for a femtocell is presented. Moreover, the user blocking probability
in a macrocell and the blocking probabilities of femtocell and macrocell
users in a femtocell are derived and analyzed. Furthermore, spectrum and energy efficiency models are proposed for
two-tier femtocell networks with partially open channels. Simulation results have shown the impacts
of critical parameters on the two-tier femtocell networks, including the number of femtocell users, the number
of femtocells in a macrocell, and the number of open or closed channels in a
femtocell. Our analysis indicates that the spectrum and energy efficiency of two-tier femtocell networks can be traded off by configuring different numbers of open channels in a femtocell. Moreover, the results of energy efficiency for two-tier femtocell networks can provide useful guidelines to determine the number of femtocells to deploy in a macrocell.


\section*{Appendix A}
The handoff probability from a femtocell into a macrocell is
defined as ${P_{Handoff\_FM}}$, which is given by \cite{Zhang10}
\[{P_{Handoff\_FM}} = \frac{{{\eta _{RT\_F}}}}{{\mu  + {\eta _{RT\_F}}}}.\tag{33}\]

The handoff probability from a macrocell into a femtocell is defined as
${P_{Handoff\_MF}}$, which is given by (34) \cite{Zhang10}.

The handoff probability from a macrocell into one of adjacent macrocells is defined as
${P_{Handoff\_MM}}$, which is given by \cite{Zhang10}
\[{P_{Handoff\_MM}} = \frac{{{\eta _{RT\_M}}}}{{\mu  + {\eta _{RT\_M}}}}.\tag{35}\]
\begin{figure*}[!t]
\[{P_{Handoff\_MF}} = \frac{{{A_F}}}{{{A_M}}} \cdot \left[ {\frac{{\ln \left( {\frac{\mu }{{{\eta _{RT\_M}}}}} \right)}}{{\frac{\mu }{{{\eta _{RT\_M}}}}}} - \frac{{{\eta _{RT\_M}}}}{\mu }\left( {{e^{ - \frac{{{\eta _{RT\_M}}}}{\mu }}} - 1} \right)} \right].\tag{34}\]
\[{\lambda _{FU\_MM}} = \left( {{\lambda _{FU\_M}} + {\lambda _{FU\_FM}} + {\lambda _{FU\_MM}}} \right) \cdot \left( {1 - {P_{U\_M}}} \right) \cdot {P_{Handoff\_MM}}.\tag{38}\]
\[{\lambda _{FU\_H}} = \frac{1}{N}\left( {{\lambda _{FU\_M}} + {\lambda _{FU\_FM}} + {\lambda _{FU\_MM}}} \right) \cdot \left( {1 - {P_{U\_M}}} \right) \cdot {P_{Handoff\_MF}},\tag{39a}\]
\[{\lambda _{FU\_M}} = NM\left( {1 - q} \right) \cdot {\lambda _F},\tag{39b}\]
\[{\lambda _{FU\_FM}} = N{\lambda _1} \cdot \left( {1 - {P_{FU\_F}}} \right) \cdot {P_{Handoff\_FM}},\tag{39c}\]
\[{\lambda _{FU\_MM}} = \left( {{\lambda _{FU\_M}} + {\lambda _{FU\_FM}} + {\lambda _{FU\_MM}}} \right) \cdot \left( {1 - {P_{U\_M}}} \right) \cdot {P_{Handoff\_MM}}.\tag{39d}\]
\[{\lambda _{MU\_MM}} = \left( {{\lambda _{MU\_M}} + {\lambda _{MU\_FM}} + {\lambda _{MU\_MM}}} \right) \cdot \left( {1 - {P_{U\_M}}} \right) \cdot {P_{Handoff\_MM}}.\tag{42}\]
\[{\lambda _{MU\_H}} = \frac{1}{N}\left( {{\lambda _{MU\_M}} + {\lambda _{MU\_FM}} + {\lambda _{MU\_MM}}} \right) \cdot \left( {1 - {P_{U\_M}}} \right) \cdot {P_{Handoff\_MF}},\tag{43a}\]
\[{\lambda _{MU\_M}} = \left( {1 - N\frac{{{A_F}}}{{{A_M}}}} \right) \cdot {\lambda _M},\tag{43b}\]
\[{\lambda _{MU\_FM}} = N{\lambda _2} \cdot \left( {1 - {P_{MU\_F}}} \right) \cdot {P_{Handoff\_FM}},\tag{43c}\]
\[{\lambda _{MU\_MM}} = \left( {{\lambda _{MU\_M}} + {\lambda _{MU\_FM}} + {\lambda _{MU\_MM}}} \right) \cdot \left( {1 - {P_{U\_M}}} \right) \cdot {P_{Handoff\_MM}}.\tag{43d}\]
\end{figure*}

Active femtocell users in a macrocell can be further divided into three types of femtocell
users: 1) a femtocell user with a new call in the specified macrocell, whose traffic
arrival rate is ${\lambda _{FU\_M}}$; 2) an active femtocell user handed off
from a femtocell into the specified macrocell, whose traffic arrival rate is ${\lambda
_{FU\_FM}}$; 3) an active femtocell user handed off from an adjacent
macrocell into the specified macrocell, whose traffic arrival rate is ${\lambda
_{FU\_MM}}$. In this case, ${\lambda _{FU\_M}}$ is expressed as
\[{\lambda _{FU\_M}} = NM\left( {1 - q} \right) \cdot {\lambda _F},\tag{36}\]
${\lambda _{FU\_FM}}$ is given by
\[{\lambda _{FU\_FM}} = N{\lambda _1} \cdot \left( {1 - {P_{FU\_F}}} \right) \cdot {P_{Handoff\_FM}}.\tag{37}\]

To keep a balance in a stationary system, the outgoing traffic rate of
femtocell users should be equal to the entering traffic rate of femtocell
users in a macrocell \cite{Lin10}. Therefore, ${\lambda _{FU\_MM}}$ is given by (38).

Furthermore, the handoff-in traffic arrival rate ${\lambda _{FU\_H}}$ is
derived as follows (39).

\section*{Appendix B}
Active macrocell users in a macrocell can be further divided into three types of macrocell
users: 1) a macrocell user with a new call in the specified macrocell, whose traffic
arrival rate is ${\lambda _{MU\_M}}$; 2) an active macrocell user handed off
from a femtocell into the specified macrocell, whose traffic arrival rate is ${\lambda
_{MU\_FM}}$; 3) an active macrocell user handed off from an adjacent
macrocell into the specified macrocell, whose traffic arrival rate is ${\lambda
_{MU\_MM}}$. In this case, ${\lambda _{MU\_M}}$ is given by
\[{\lambda _{MU\_M}} = \left( {1 - N\frac{{{A_F}}}{{{A_M}}}} \right) \cdot {\lambda _M}.\tag{40}\]
${\lambda _{MU\_FM}}$ is given by
\[{\lambda _{MU\_FM}} = N{\lambda _2} \cdot \left( {1 - {P_{MU\_F}}} \right) \cdot {P_{Handoff\_FM}}.\tag{41}\]

To keep a balance in a stationary system, the leaving traffic rate of
macrocell users should be equal to the entering traffic rate of macrocell
users in a macrocell \cite{Lin10}. Therefore, ${\lambda _{MU\_MM}}$ is given by (42).

Furthermore, the handoff traffic arrival rate of macrocell users ${\lambda _{MU\_H}}$ is
derived by (43).


\end{document}